\documentclass[conference,a4paper]{IEEEtran}
\usepackage{graphicx}
\usepackage{amsmath}
\usepackage{amssymb}
\usepackage{booktabs}
\usepackage[hyphens]{url}
\usepackage[colorlinks=true,allcolors=blue]{hyperref}
\usepackage{algpseudocode}
\usepackage{algorithm}
\usepackage{tabularx}
\usepackage{multirow}
\usepackage{float}
\usepackage{orcidlink}

\title{Approximate Convex Decomposition (ACD) with Divide and Conquer Strategy for DRL-Based BS Localization}

\author{
    \IEEEauthorblockN{Omar Awad $^1$\orcidlink{}, Aly Morsy $^1$\orcidlink{}, Mohamed Shalma $^2$\orcidlink{0000-0002-7185-1393}}
    \IEEEauthorblockA{$^1$ Faculty of Media Engineering and Technology, The German University in Cairo\\
    $^2$ Faculty of Information Engineering and Technology, The German University in Cairo\\
    }
}

\begin{document}

\maketitle

\begin{abstract}
    
Optimal base station (BS) placement is a fundamental challenge in wireless localization. Conventional optimization approaches may become computationally demanding as the geometric complexity of the environment increases, limiting their scalability to realistic building layouts. This paper proposes an Approximate Convex Decomposition (ACD)-based divide-and-conquer framework integrated with deep reinforcement learning (DRL) for optimizing BS placement. The proposed approach exploits the geometric structure of building layouts by partitioning the original environment into smaller sub-regions using cuts associated with reflex vertices. Furthermore, the proposed solution is bench-marked against DRL without ACD and with the optimal solution obtained by exhaustive search. Results show the superiority of the proposed approach and that it approaches the optimal solution.

\end{abstract}


\section{Introduction}
Artificial intelligence (AI) and computer science (CS) are driving advances in intelligent decision-making, data-driven optimization, and automated problem-solving across engineering applications. In wireless communications, these techniques enable intelligent network planning by learning effective solutions to complex optimization problems. Base station (BS) placement is a key challenge in 6G networks \cite{8959607}, particularly in buildings with irregular, non-convex geometries and nonuniform user distributions. Deep reinforcement learning (DRL) offers a data-driven approach to placement optimization. Recent works have shown great potentials of DRL with respect to traditional techniques \cite{10174755,11232082,8761276,11313425}. However, exploring complex spatial search spaces can be challenging especially from mobile operators perspective in real non-convex buildings topologies.

For optimal BS placement, the related works, specifically in terms of handling non-convex topologies remain limited. \cite{9149258} addressed non-uniform wireless coverage caused by building blockage using a two-stage DRL framework, where a preliminary placement based on a dominant-LoS model is refined using site-specific LoS/NLoS information extracted from a 3D terrain map. However, their approach does not explicitly exploit the geometric structure of non-convex building floor plans to decompose the BS placement search space. In \cite{perner2026multiagentreinforcementlearningbase}, the authors proposed a ray-tracing-assisted multi-agent reinforcement learning (MARL) framework for TDOA-based BS placement, training PPO agents on channel impulse responses generated from a 3D university campus model to account for NLOS propagation and multipath-induced localization errors. Their approach optimizes BS configurations over a discrete candidate grid but does not explicitly exploit the geometric structure of non-convex building floor plans to reduce the placement search space. The authors in \cite{heydari2021reinforcementlearningbasedcoveragepath} proposed a reinforcement learning (RL)-based coverage path planning framework that improves area coverage and reduces overlap by learning navigation policies, including a hybrid approach that combines zigzag movement with RL to overcome the limitations of greedy repositioning. However, their method primarily addresses robot navigation and coverage in grid-based environments, rather than exploiting the geometric structure of non-convex environments for wireless BS placement optimization. Finally, Evolutionary optimization has also been investigated for BS deployment in dense urban 5G networks. In \cite{555}, the authors developed a GA-based framework incorporating adaptive selection, crossover, mutation, and local search to jointly optimize the number and locations of BSs. Their formulation considers received signal power, SINR, intersystem interference, user data rates, and QoS constraints, while penalizing overloaded BSs and excessive deployment density. Yet, the non-convex nature of the BS vicinity remains unexplored due to the extreme hardness of the NP-hard optimization problem. On the other hand, the promising advantages of ACD has been recently shown in different applications including robotics and computer science \cite{11164152,10.1145/3528223.3530103}.

In contrast to the above-mentioned BS placement related works, this work leverages approximate convex decomposition (ACD) to partition non-convex building topologies at reflex vertices into smaller subregions. The contributions of this work can be summarized as:
\begin{itemize}
\item A non-convex NP-hard BS localization problem is proposed using realistic non-convex building topology.
\item To deal with this non-convexity, an approximate convex decomposition (ACD) framework is proposed using the divide and conquer startegy to decompose the original problem into smaller convex ones.
\item The proposed framework is bench-marked against DRL frameworks without ACD and with the optimal solution obtained by exhaustive search.
\end{itemize}

\section{System Model}
\label{sec:System}
\begin{figure*}[t]
\centering
\includegraphics[width=0.75\textwidth]{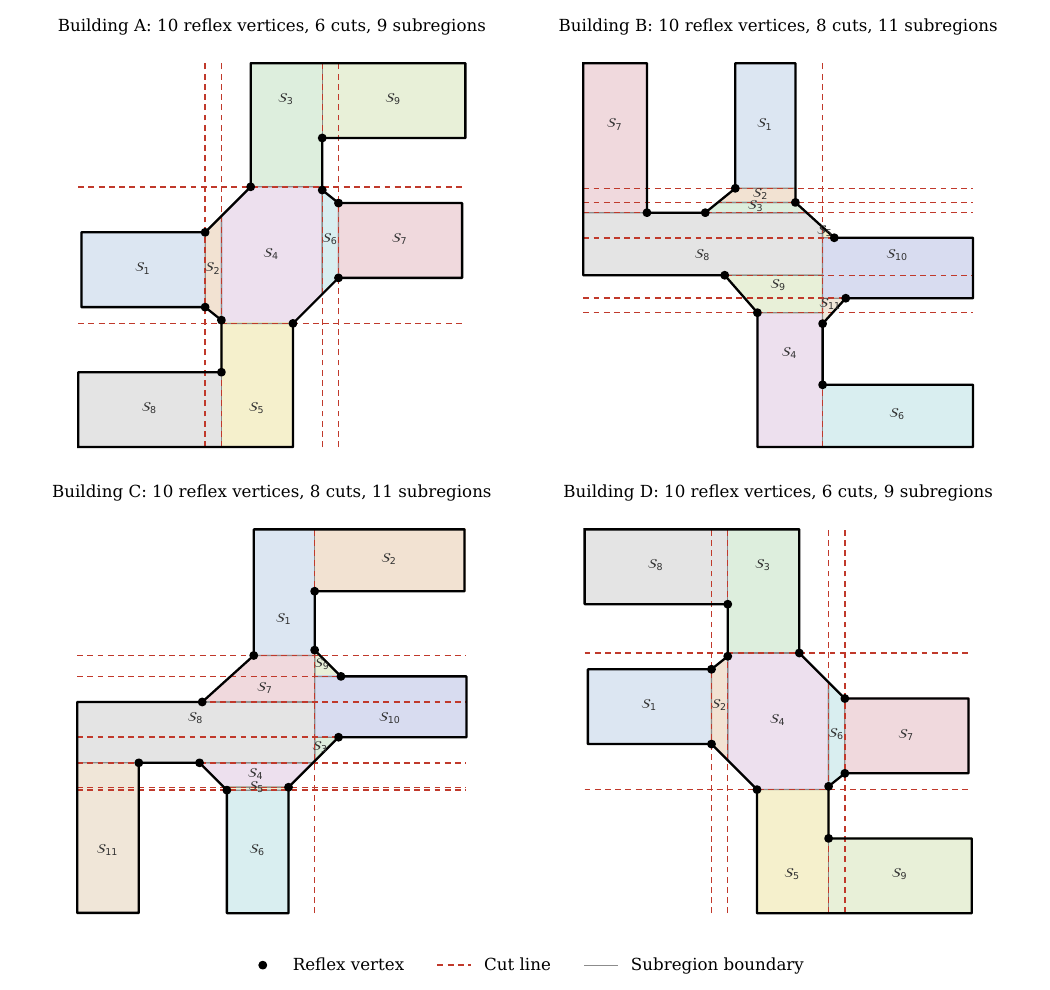}
\caption{ACD of the four campus buildings: reflex vertices (dots), derived cut lines (dashed) and the resulting convex subregions $\mathcal{S}_j$.}
\label{fig:acd}
\end{figure*}

We consider a wireless network comprising four BSs deployed within the coverage area of four irregular building layouts labeled from (A–D) as shown in Fig. \ref{fig:acd}, each serving 500 users distributed nonuniformly across the area, with 60\% concentrated in six crowded spots of approximately 30 m across and the remaining 40\% randomly scattered. The set of feasible BS positions is the building footprint, a simple polygon $\mathcal{B}\subset\mathbb{R}^2$. Campus buildings typically consist of several wings joined at concave corners, which makes $\mathcal{B}$ non-convex. Each building has 10 reflex vertices and is partitioned using approximate convex decomposition (ACD) into 9 or 11 subregions through 6 or 8 cuts, respectively. The objective is to determine the BS position that maximizes the worst achievable rate among all users to maximize fairness. The proposed divide-and-conquer approach trains a DRL soft actor–critic (SAC) agent within each subregion and selects the best-performing candidate position across the subregions, while a whole-building agent serves as a benchmark. The optimization is evaluated using the same user locations across three independent training seeds to ensure a consistent comparison. The geometric center of each building's bounding box is also considered as a reference placement to quantify the impact of nonuniform user distributions on the optimal BS location. Performance is assessed based on the achievable rate and its gap from the best possible rate for each building that is obtained using exhaustive search.

\section{Proposed Approximate Convex Decomposition of the non-Convex Topology}
\label{sec:acd}

The proposed strategy decomposes $\mathcal{B}$ into convex subregions by approximate convex decomposition (ACD), solves the placement problem independently inside each subregion, and retains the best of the resulting solutions. This section details the decomposition algorithm, its application to the campus buildings, and the way the subregions are used for placement.

\subsection{Definitions}
Let the footprint be described by its vertices $v_1,\dots,v_n$, with $v_i=(x_i,y_i)$, listed in order along the boundary. A convex partition of $\mathcal{B}$ is a set of polygons $\{\mathcal{S}_1,\dots,\mathcal{S}_N\}$ satisfying
\begin{equation}
\mathcal{B}=\bigcup_{j=1}^{N}\mathcal{S}_j, \qquad \operatorname{int}\mathcal{S}_i\cap\operatorname{int}\mathcal{S}_j=\emptyset \quad (i\neq j),
\label{eq:partition}
\end{equation}
in which every $\mathcal{S}_j$ is convex and $\operatorname{int}$ denotes the interior of a set. ACD relaxes exact convexity to a tolerance. The concavity of a polygon $\mathcal{P}$ is measured as
\begin{equation}
\kappa(\mathcal{P}) = A\big(\operatorname{conv}\mathcal{P}\big) - A(\mathcal{P}),
\label{eq:concavity}
\end{equation}
where $A(\cdot)$ denotes area and $\operatorname{conv}\mathcal{P}$ is the convex hull of $\mathcal{P}$, and $\mathcal{P}$ is accepted as convex if $\kappa(\mathcal{P})\le\varepsilon$. In this work $\varepsilon=10^{-3}$~m$^2$, so the resulting subregions are convex up to numerical precision.

\subsection{Reflex Vertex Detection}
The non-convexity of a simple polygon is caused by its reflex vertices, i.e., the vertices whose interior angle exceeds $\pi$. For vertex $v_i$, let $e_i^{-}=v_i-v_{i-1}$ and $e_i^{+}=v_{i+1}-v_i$ be the incoming and outgoing edge vectors, with indices taken modulo $n$. The turn direction at $v_i$ is given by the $z$-component of their cross product,
\begin{equation}
c_i = e^{-}_{i,x}\,e^{+}_{i,y} - e^{-}_{i,y}\,e^{+}_{i,x}.
\label{eq:cross}
\end{equation}
Letting $\sigma=+1$ if the vertices are listed counter-clockwise and $\sigma=-1$ if they are listed clockwise, the set of reflex vertices is
\begin{equation}
\mathcal{R}=\{\,v_i : \sigma\, c_i<0 \,\}.
\label{eq:reflex}
\end{equation}
The orientation factor $\sigma$ makes the test independent of the order in which the vertices are specified. A footprint with $\mathcal{R}=\emptyset$ is already convex and is used as a single subregion.

\subsection{Cut-Line Derivation}
Each reflex vertex is resolved by an axis-parallel cut line through it. Since building footprints are dominated by axis-aligned walls, reflex vertices frequently share a coordinate, in which case a single cut resolves several of them. The set of cuts is derived from $\mathcal{R}$ in three stages:
\begin{enumerate}
\item Horizontal cuts: a horizontal cut $y=y_0$ is added for every coordinate $y_0$ shared by at least two reflex vertices,
\begin{equation}
\mathcal{H}=\big\{\,y_0 : \left|\{v\in\mathcal{R} : v_y=y_0\}\right|\ge 2 \,\big\},
\label{eq:hcuts}
\end{equation}
and the corresponding vertices are marked as resolved.
\item Vertical cuts: among the reflex vertices that remain unresolved, a vertical cut $x=x_0$ is added for every coordinate $x_0$ shared by at least two of them, and these vertices are marked as resolved.
\item Fallback cuts: each reflex vertex that is still unresolved receives its own horizontal cut, so that every reflex vertex is resolved.
\end{enumerate}
Coordinates are compared with a tolerance of $10^{-6}$~m. No cut is specified manually; all cuts follow from the geometry of the footprint.

\subsection{Cutting, Cleaning and Merging}
The cuts are applied sequentially, starting from the single piece $\mathcal{B}$. A horizontal cut $y=y_0$ splits every current piece $\mathcal{P}$ whose vertical extent strictly contains $y_0$ into $\mathcal{P}\cap\{y\ge y_0\}$ and $\mathcal{P}\cap\{y\le y_0\}$, whereas pieces not crossed by the line remain unchanged; vertical cuts act analogously on the $x$-coordinate. Because every piece results from an intersection with the footprint, areas outside the building, such as the space between two wings, are excluded automatically. Intersections producing multi-part geometries are split into their polygonal components, and slivers with an area below 1~m$^2$, which arise when a cut grazes a vertex, are discarded.

Since each cut extends across every piece it crosses, the cutting stage over-segments the footprint. The pieces are therefore merged greedily: whenever two pieces are adjacent, i.e., their boundaries intersect while their interiors do not, and their union satisfies $\kappa\le\varepsilon$ according to \eqref{eq:concavity}, they are replaced by their union. This step is repeated until no pair of adjacent pieces can be merged. Finally, each subregion is verified against \eqref{eq:concavity}, and the total area of the subregions is compared with $A(\mathcal{B})$ to confirm that \eqref{eq:partition} holds.

\subsection{Application to the Buildings}
Table~\ref{tab:acd} summarizes the decomposition of the four campus buildings, and Fig.~\ref{fig:acd} shows the detected reflex vertices, the derived cuts and the resulting subregions. Each footprint has 20 vertices, 10 of which are reflex. The cutting stage yields 11 pieces for buildings A and D and 17 pieces for buildings B and C, which the merging stage reduces to 9 and 11 convex subregions, respectively. For all four buildings every subregion satisfies $\kappa\le\varepsilon$, with the largest observed concavity below $10^{-12}$~m$^2$, and the subregion areas sum to the footprint area. The longest side of a subregion ranges from 32 to 44~m for buildings A and D and from 3 to 68~m for buildings B and C, compared with footprint extents of 110 to 119~m. The smallest subregions are narrow strips created where cuts pass close to chamfered corners, the smallest one having an area of 4.9~m$^2$ in building B.

\begin{table}[t]
\centering
\caption{ACD of the Campus Buildings ($n$: vertices, $|\mathcal{R}|$: reflex vertices, H/V: horizontal/vertical cuts, $N_{\text{cut}}$: pieces after cutting, $N$: subregions after merging)}
\label{tab:acd}
\begin{tabular}{@{}lccccc@{}}
\toprule
Building & $n$ & $|\mathcal{R}|$ & Cuts (H/V) & $N_{\text{cut}}$ & $N$ \\
\midrule
A & 20 & 10 & 6 (2/4) & 11 & 9 \\
B & 20 & 10 & 8 (7/1) & 17 & 11 \\
C & 20 & 10 & 8 (7/1) & 17 & 11 \\
D & 20 & 10 & 6 (2/4) & 11 & 9 \\
\bottomrule
\end{tabular}
\end{table}

\subsection{Placement over the Subregions}
Let $u_k=(x_k,y_k)$, $k=1,\dots,K$, denote the horizontal positions of the users, and let $p=(x,y)\in\mathcal{B}$ denote the horizontal position of the BS. With the BS and user antennas at heights $h_0$ and $h_1$, the BS--user distance is
\begin{equation}
d_k(p)=\sqrt{\lVert p-u_k\rVert^2+(h_0-h_1)^2},
\label{eq:dist}
\end{equation}
and the rate of user $k$ is $r\big(d_k(p)\big)$, where $r(\cdot)$ is determined by the path-loss model and decreases monotonically with distance. The placement objective is the rate of the worst-served user,
\begin{equation}
F(p)=\min_{k}\ r\big(d_k(p)\big).
\label{eq:objective}
\end{equation}
Each subregion $\mathcal{S}_j$ is assigned an independent DRL agent, which searches for a position $p_j\in\mathcal{S}_j$ maximizing $F$ over all $K$ users; the agents do not exchange information. The BS is then placed in the subregion whose agent achieved the highest objective,
\begin{equation}
j^{\star}=\arg\max_{j\in\{1,\dots,N\}} F(p_j), \qquad p^{\star}=p_{j^{\star}}.
\label{eq:select}
\end{equation}
Since the subregions cover $\mathcal{B}$, the best of the subregion optima coincides with the optimum over the whole footprint.

The decomposition simplifies the task of each agent in two ways. First, because $r(\cdot)$ is decreasing, maximizing $F$ is equivalent to minimizing the distance to the farthest user, $g(p)=\max_k d_k(p)$. Each $d_k$ is a convex function of $p$, hence $g$ is convex, and within a convex subregion every local minimum of $g$ is a global one: each agent faces a search problem with a single optimum, whereas over the non-convex footprint several local optima may exist. Second, each agent searches only its own subregion, whose extent is a fraction of that of the whole footprint. For an agent whose output is scaled to the extent of its search region, a given precision of the output therefore translates into a proportionally finer placement precision in metres. The complete procedure is summarized in Algorithm~\ref{alg:acd}.

\algrenewcommand\algorithmicrequire{\textbf{Input:}}
\algrenewcommand\algorithmicensure{\textbf{Output:}}
\begin{algorithm}[t]
\caption{ACD Divide-and-Conquer BS Placement}
\label{alg:acd}
\begin{algorithmic}[1]
\Require footprint $\mathcal{B}$ with vertices $v_1,\dots,v_n$; user positions $u_1,\dots,u_K$; tolerance $\varepsilon$; training steps $T$ per agent
\Ensure BS position $p^{\star}$
\State $\mathcal{R}\gets\{v_i : \sigma c_i<0\}$ \Comment{reflex vertices, \eqref{eq:reflex}}
\If{$\mathcal{R}=\emptyset$}
    \State $\mathcal{S}\gets\{\mathcal{B}\}$ \Comment{already convex}
\Else
    \State derive the cut set $\mathcal{C}$ from $\mathcal{R}$ (horizontal, vertical and fallback cuts)
    \State $\mathcal{S}\gets\{\mathcal{B}\}$
    \For{each cut $\ell\in\mathcal{C}$}
        \State split every piece of $\mathcal{S}$ crossed by $\ell$ into its two sides
    \EndFor
    \State discard pieces with an area below 1~m$^2$
    \While{adjacent $\mathcal{P}_a,\mathcal{P}_b\in\mathcal{S}$ with $\kappa(\mathcal{P}_a\cup\mathcal{P}_b)\le\varepsilon$ exist}
        \State $\mathcal{S}\gets\big(\mathcal{S}\setminus\{\mathcal{P}_a,\mathcal{P}_b\}\big)\cup\{\mathcal{P}_a\cup\mathcal{P}_b\}$
    \EndWhile
\EndIf
\For{$j=1,\dots,N$} \Comment{$N=|\mathcal{S}|$ independent agents}
    \State train a SAC agent on $\mathcal{S}_j$ for $T$ steps with reward $F(p)$, $p\in\mathcal{S}_j$
    \State $p_j\gets$ position selected by the trained agent
\EndFor
\State $j^{\star}\gets\arg\max_{j} F(p_j)$ \Comment{\eqref{eq:select}}
\State \Return $p^{\star}=p_{j^{\star}}$
\end{algorithmic}
\end{algorithm}

\begin{figure*}
    \centering
    \includegraphics[width=0.75\linewidth]{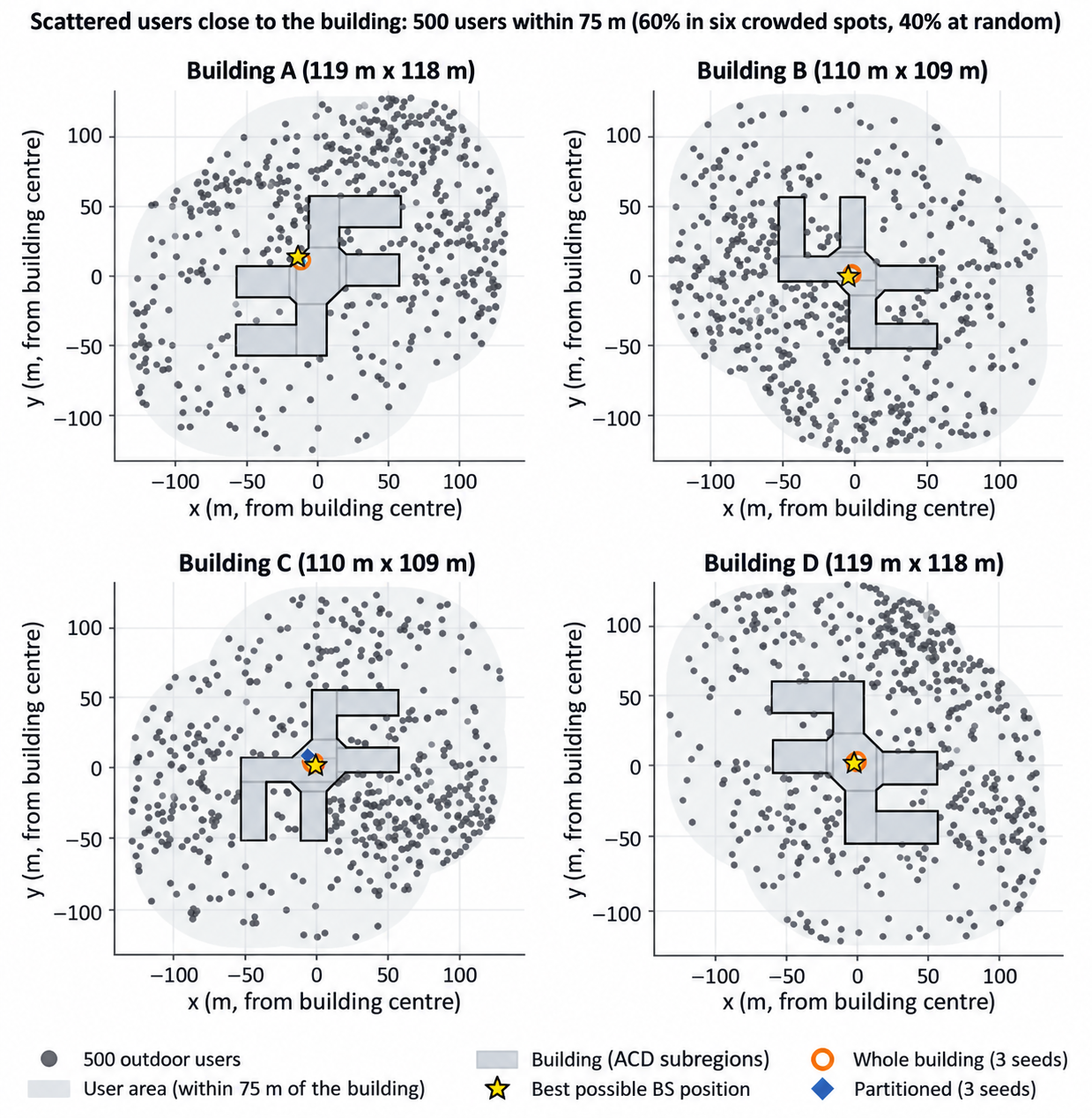}
    \caption{Spatial distributions of the 500 users across the four non-convex building layouts (A–D), illustrating the six crowded regions and the randomly distributed users, together with the corresponding reference-optimal BS placement positions. }
    \label{fig:Rs1}
\end{figure*}
\begin{figure*}
    \centering
    \includegraphics[width=\linewidth]{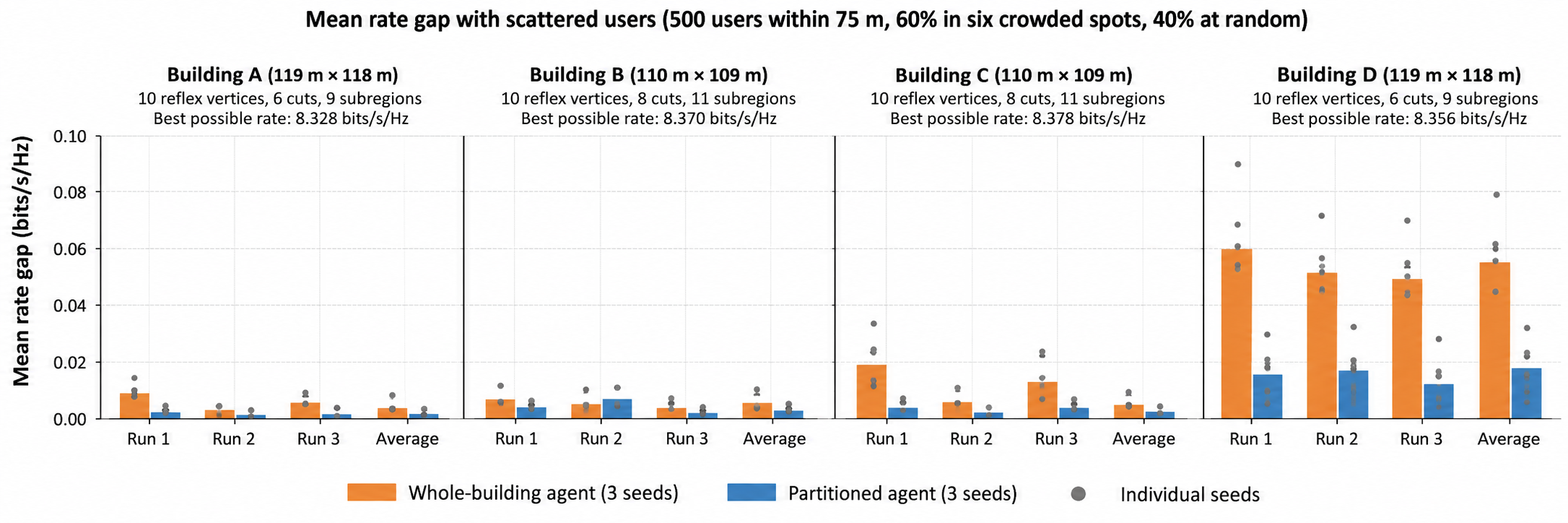}
    \caption{Performance comparison of the proposed ACD-based divide-and-conquer SAC approach and the whole-building SAC baseline across the four building layouts. The results illustrate the achievable-rate gaps relative to the reference best rates under nonuniform user distributions.}
    \label{fig:Rs2}
\end{figure*}
\section{Numerical Results}

The proposed ACD-based partitioned DRL approach is evaluated using the four building layouts introduced in Section II. For each building, 500 users are placed within the coverage area, with 60\% concentrated in six crowded regions of approximately 30 m across and 40\% distributed randomly. The same user locations are retained across all three random training seeds, while the SAC training procedure uses 20,000 steps per agent.

Fig. \ref{fig:Rs1} illustrates the 500 scattered users for the four building layouts and the corresponding best possible BS positions. The best possible rates range from 8.328 to 8.378 bits/s/Hz across the four layouts. More importantly, the optimal BS position is displaced from the building center by 2 m, 14 m, 16 m, and 9 m for buildings A, B, C, and D, respectively. These results demonstrate that the geometric center alone is not a reliable indicator of the best deployment position when user locations are spatially nonuniform.

For benchmarks, the center-based placement i.e., placing each BS exactly in the center of the building exhibits rate gaps of 0.015, 0.013, 0.027, and 0.055 bits/s/Hz from the optimal solution for buildings A–D, respectively. The largest loss occurs in building D, highlighting the potential performance penalty of adopting a fixed geometric placement without accounting for the user distribution. In contrast, both DRL-based methods learn placement decisions that approach the best possible rate.

\begin{table}[t]
\centering
\caption{ Rate-Gap Comparison Across the Four Building Layouts}
\label{tab:mean_gap}
\begin{tabular}{lccc}
\toprule
Building & Partitioned SAC & Whole-Building SAC & Gap Reduction (\%) \\
\midrule
A & 0.0004 & 0.0021 & 81.0 \\
B & 0.0016 & 0.0016 & 0.0 \\
C & 0.0005 & 0.0015 & 66.7 \\
D & 0.0003 & 0.0028 & 89.3 \\
\midrule
Average & 0.0007 & 0.0020 & 65.0 \\
\bottomrule
\end{tabular}
\end{table}

The partitioned approach achieves a smaller mean rate gap than the whole-building agent in buildings A, C, and D, while both approaches exhibit the same mean gap in building B. Specifically, the partitioned method achieves mean gaps of 0.0004, 0.0016, 0.0005, and 0.0003 bits/s/Hz for buildings A–D, respectively. The corresponding whole-building gaps are 0.0021, 0.0016, 0.0015, and 0.0028 bits/s/Hz which are summarized in TABLE \ref{tab:mean_gap}. For buildings A, C, and D, the partitioned method reduces the mean rate gap by approximately 81.0\%, 66.7\%, and 89.3\%, respectively, relative to the whole-building agent. In building B, both approaches attain the same mean gap of 0.0016 bits/s/Hz. These findings indicate that geometry-aware partitioning can improve placement accuracy in several building configurations, although the benefit depends on the topology and the resulting optimization landscape.

The maximum gaps across the three seeds further illustrate the consistency of the partitioned approach. Its worst-seed gaps are 0.0008, 0.0021, 0.0008, and 0.0005 bits/s/Hz for buildings A–D, respectively, compared with 0.0040, 0.0019, 0.0032, and 0.0037 bits/s/Hz for the whole-building agent. Thus, the partitioned approach also exhibits smaller worst-seed gaps in buildings A, C, and D, whereas building B shows a slightly larger worst-seed gap despite having the same mean gap.

Overall, the results demonstrate that combining ACD with SAC-based placement optimization yields consistently accurate BS placement across four geometrically distinct building layouts and a nonuniform user distribution. The partitioned approach attains the same mean performance as the whole-building agent in one layout and smaller mean rate gaps in the other three. These findings support the potential of geometry-aware decomposition as a practical means of structuring DRL-based BS placement in complex indoor environments.

\section{Conclusion}
This paper proposed an ACD-based divide-and-conquer framework for DRL-based BS placement in non-convex building environments. By decomposing complex building layouts into smaller convex subregions, independent SAC agents optimize the placement problem within reduced search spaces. Numerical results across four building layouts demonstrate an average rate-gap reduction of 65\% compared with whole-building SAC, with improvements of up to 89.3\%, while closely approaching the reference optimal solution. These findings highlight the effectiveness of geometry-aware decomposition in improving DRL-based BS placement in complex environments.

\bibliographystyle{IEEEtran}
\bibliography{Ref}

\end{document}